\documentclass{aa}
\usepackage{graphicx}
\usepackage{txfonts}
\usepackage{natbib}
\begin{document}
\title{OH/IR stars and their superwinds as observed by the Herschel
Space Observatory
\thanks{Herschel is an ESA space observatory with science instruments 
provided by European-led Principal Investigator consortia and with 
important participation from NASA.}
}
\author{K. Justtanont\inst{\ref{inst1}}
        \and D. Teyssier\inst{\ref{inst2}}
        \and M.J. Barlow\inst{\ref{inst3}}
        \and M. Matsuura\inst{\ref{inst3}}
        \and B. Swinyard\inst{\ref{inst3},\ref{inst4}}
        \and L.B.F.M. Waters\inst{\ref{inst5},\ref{inst6}}
        \and J. Yates\inst{\ref{inst3}}
}
\institute{Chalmers University of Technology, Onsala Space Observatory,
           S-439 92 Onsala, Sweden\label{inst1}
\and European Space Astronomy Centre, ESA, P.O. Box 78, E-28691
Villanueva de la Ca\~nada, Madrid, Spain\label{inst2}
\and University College London, Dept. of Physics \& Astronomy, Gower Street,
     London, WC1E 6BT, UK\label{inst3}
\and Space Science and Technology Department, Rutherford Appleton Laboratory, 
     Oxfordshire OX11 0QX, UK\label{inst4}
\and SRON Netherlands Institute for Space Research, Sorbonnelaan 2, 
     3584 CA Utrecht, The Netherlands\label{inst5}
\and Sterrenkundig Instituut Anton Pannekoek, Universiteit van Amsterdam,
     Postbus 94249, 1090 GE Amsterdam, The Netherlands\label{inst6}
}
\date{1 May 2013 / 7 June 2013}
\abstract
{}
{
In order to study the history of mass loss in extreme OH/IR stars,
we observed a number of these objects using CO as a tracer of the 
density and temperature structure of their circumstellar envelopes.}
{
Combining CO observations from the Herschel Space Observatory with
those from the ground, we trace mass loss rates 
as a function of radius in five extreme OH/IR stars. 
Using radiative transfer modelling, we modelled the dusty envelope
as well as the CO emission. The high-rotational transitions of CO
indicate that they originate in a dense superwind region close to
the star while the lower
transitions tend to come from a more tenuous outer wind
which is a result of the mass loss since the early AGB phase.}
{
The models of the circumstellar envelopes around these stars suggest
that they have entered a superwind phase in the past
200 - 500 years. The low $^{18}$O/$^{17}$O ($\sim$ 0.1 
compared to the solar abundance ratio of $\sim$ 5) and $^{12}$C/$^{13}$C
(3-30 cf. the solar value of 89) 
ratios derived from our study support the idea that 
these objects have undergone hot-bottom burning and hence that they
are massive M $\geq$ 5 M$_{\odot}$ AGB stars. }
{}
{
\maketitle

\section{Introduction}

A class of asymptotic giant branch (AGB) stars known as extreme
OH/IR stars are among the objects which 
underdo intensive mass loss. They were originally identified by 
the presence of strong
1612~MHz OH masers \citep[e.g.,][]{johansson77, telintel89, sevenster02}.
Later IRAS LRS survey found 
that these stars have highly dusty circumstellar envelopes
with the 10 and 20~$\mu$m silicate features both in absorption.
Although dust features of OH/IR stars indicate a very high mass loss rate,
ground-based CO observations of some
of these objects showed very weak emission \citep{heske90} which implied
that the mass loss rates were a couple of order of magnitudes lower than those 
derived from the dust.
A detailed study of OH~26.5+0.6 revealed that the star has recently
increased its mass loss rate in the past $\sim$ 200 years 
\citep{justtanont96}. This sudden increase in mass loss rate is called
a superwind \citep{iben83}. Theoretical calculations 
show that the superwind phase occurs towards the end of the AGB evolution
\citep[e.g.,][]{vassiliadis93} when most of the initial mass of the
star is lost, allowing the central star to evolve towards a planetary
nebula phase.
\citet{delfosse97} observed low $^{12}$CO/$^{13}$CO
ratios in this class of objects and concluded that the central stars underwent 
hot-bottom burning to achieve such a ratio. 

The isotopic ratios of AGB stars are keys to determining the stellar 
mass and the subsequent evolution, because the AGB nucleosynthesis 
processes depend heavily on their initial masses.
Observations by \citet{smith90} of several O-rich AGB stars in the 
Magellanic Clouds showed unexpectedly large 
lithium abundances with a correlation between the abundance and the 
pulsation period. \cite{sackmann92} predicted a high lithium abundance of 
log~$\epsilon (^{7}$Li) $\sim$ 4-4.5, 
an order of magnitude higher than the cosmic value,
for a star which has undergone hot-bottom burning at the base of its
convective envelope. The temperature in this layer is 5\,10$^{6}$ K and is
reached by stars of mass $\textgreater$ 4 M$_{\odot}$. The Li abundance in 
these models depended on the initial metallicity. 
However, Li is easily destroyed as it passes through the convective
envelope. A few
carbon-rich AGB stars have also been observed to have high Li abundances.
These are probably AGB stars with an initial mass of $\sim$ 4 M$_{\odot}$ 
which have begun hot-bottom burning. Both the carbon and oxygen isotopes
are affected by this process -- 
$^{13}$C is produced from $^{12}$C, thereby reducing the
$^{12}$C abundance throughout the envelope, which results in a lowering
of the C/O ratio. This can lead carbon-rich AGB stars 
to ultimately become oxygen-rich again. 
Also, $^{18}$O is destroyed while $^{17}$O remains unchanged
relative to $^{16}$O during this process.
The phase with a low $^{12}$C/$^{13}$C ratio can last $\sim 10^{5}$
yr for a 5-6 M${_\odot}$ star \citep{lattanzio03}.

Many of the envelopes around these extreme OH/IR stars exhibit the presence of
water-ice \citep[e.g.][]{sylvester99,dijkstra03, maldoni03, justtanont06} 
as well as
crystalline silicates \citep[e.g.][]{cami98, sylvester99, molster02, suh02}.
The dust temperature is expected to be below 100-200 K for amorphous 
and crystalline water-ice to condense out onto existing
silicate grains \citep{hudgins93} 
which requires that the envelope be highly self-shielded,
i.e., the density is high. The formation of crystalline silicates requires
heating of the amorphous silicate \citep{kozasa99}
which can occur in the innermost part
of the envelope due to both dust-drag and radiative back-heating of the
grains themselves.

In this paper, we present observations of CO rotational lines 
from a number of extreme OH/IR stars 
observed by the {\it Herschel Space Observatory}
\citep{pilbratt10} in order to study the temporal behaviour of the mass loss in
these objects,
as well as of isotopologues of H$_{2}$O which can be used as initial mass
indicators.
Section 2 describes the observations and the results
of the radiative transfer modelling are presented in section 3 for 
individual objects. The implications of the isotopologues of CO and
H$_{2}$O and the superwind are discussed in section 4 and the 
results are summarized in section 5.

\section{Observations}

\begin{figure*}[t]
\centering
\includegraphics[width=17cm]{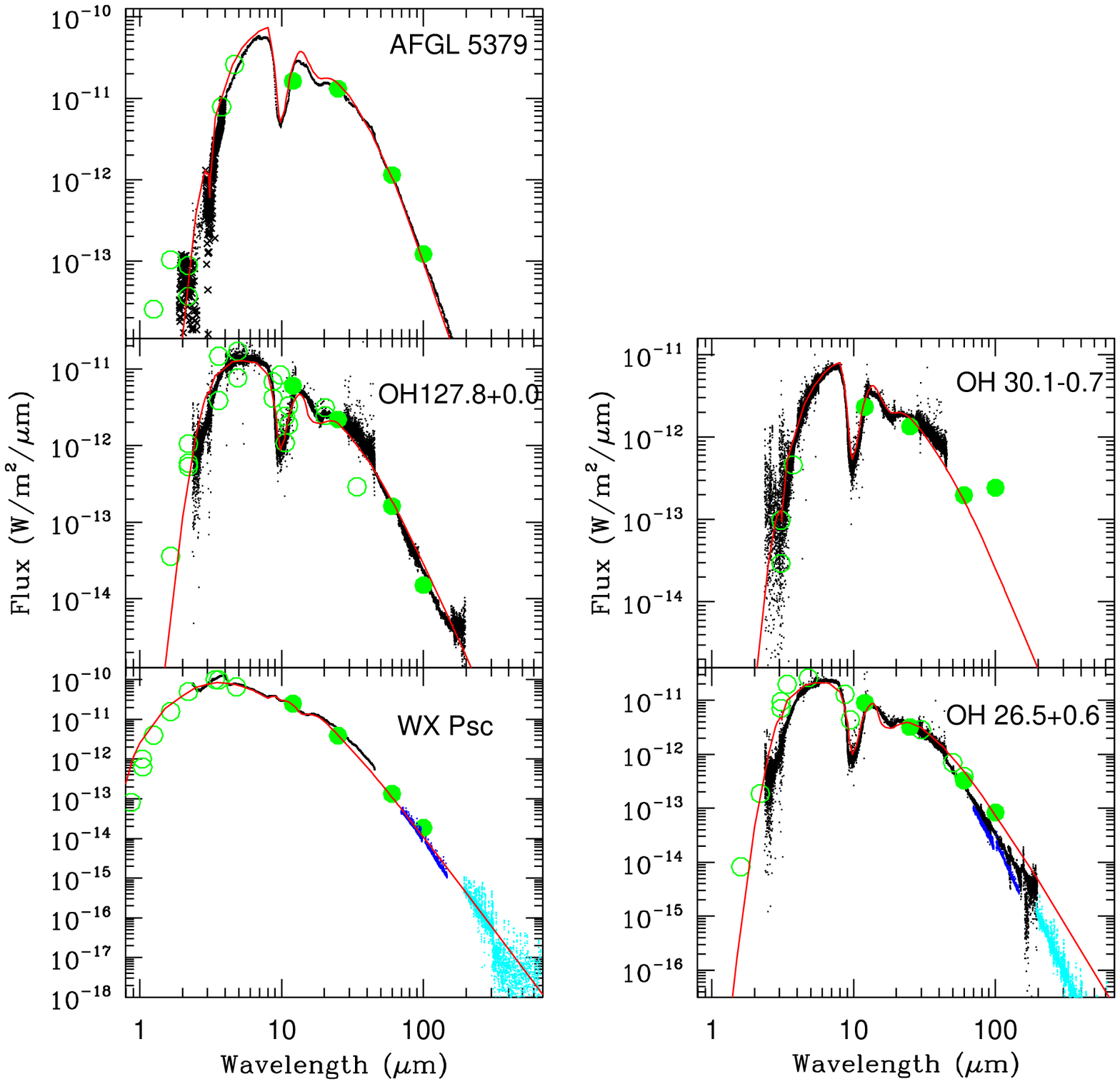}
\caption{SED fits for the sample stars (solid line) to the ISO spectra
(black dots) and photometric points from IRAS (filled green circles).
The PACS (blue dots) and SPIRE (cyan dots) spectra are also plotted when
available.
The published photometric data (open circles) are taken from \citet{dyck74} and
\citet{epchtein80} for WX Psc, \citet{persi90} for OH~127.8+0.0, 
\citet{garcia97} and \citet{lepine95} for AFGL~5379, \citet{werner80}
for OH~26.5+0.6, and \citet{justtanont06} for both OH~26.5+0.6 and OH~30.1-0.7.
}
\label{fig-sed}
\end{figure*}

We obtained spectrally resolved observations of the CO lines using the 
heterodyne instrument
\citep[HIFI,][]{degraauw10} aboard {\it Herschel}
from the guaranteed time programs HIFISTARS (P.I. V. Bujarrabal)
and SUCCESS (P.I. D. Teyssier) which observed
a total of 5 OH/IR stars in different combinations of CO J=5-4, 6-5, 9-8, 
10-9 and 16-15 (Table~\ref{tab-sources}). 
The HIFI data were averaged over both H- and V-polarizations and
corrected for the beam efficiency \citep{roelfsema12},
as well as being corrected baseline distortion by subtracting
a polynomial fit to the spectra. A full description of how the data are treated
can be found in \citet{justtanont12}.
The high-J transitions probe the warm molecular envelope
close to the wind acceleration zone where dust grains impart their
momentum to the gas and drive the wind to the observed terminal velocity.
Observations of the three most abundant isotopologues of H$_{2}$O were
obtained as part of the HIFISTARS program. In this paper, we present
recently obtained data for AFGL~5379 and OH~26.5+0.6 for the transition
3$_{12} - 3_{03}$.

\begin{table*}[h]
\caption{{\it Herschel} observations of objects in this study
with the observing time (t$_{\rm obs}$) in seconds.}
\label{tab-sources}
\begin{tabular}{llllll}
\hline \hline
name & RA (2000) & Dec (2000) & Obs ID & observation & t$_{\rm obs}$ (s)\\
\hline
WX Psc      & 01 06 25.98 & +12 35 53.0 & 1342200967 & CO 6-5   &617\\
            &             &             & 1342201116 & CO 10-9  &1618\\
            &             &             & 1342201665 & CO 16-15 &3207 \\
            &             &             & 1342246973 & SPIRE FTS&2530 \\
            &             &             & 1342202121 & PACS Spec-red &2263 \\
            &             &             & 1342202122 & PACS Spec-blue&1135 \\
OH~127.8+0.0 & 01 33 51.21 & +62 26 53.2 & 1342201529 & CO 5-4   &134\\
            &             &             & 1342211357 & CO 9-8   &921\\
AFGL~5379   & 17 44 24.01 & -31 55 35.5 & 1342192553 & CO 6-5   &617 \\
            &             &             & 1342194741 & CO 10-9  &1618\\
            &             &             & 1342214491 & CO 16-15 &2992 \\
            &             &             & 1342195079 & 
                                     H$_{2}$O 1$_{11} - 0_{00}$ &1487 \\
            &             &             & 1342250605 & 
                                     H$_{2}$O 3$_{12} - 3_{03}$ &3160 \\
OH~26.5+0.6  & 18 37 32.51 & -05 23 59.2 & 1342194557 & CO 6-5   &617\\
            &             &             & 1342195079 & CO 10-9  &1618\\
            &             &             & 1342194774 & CO 16-15 &2992 \\
            &             &             & 1342206388 & 
                                     H$_{2}$O 1$_{11} - 0_{00}$ &1487 \\
            &             &             & 1342244511 & 
                                     H$_{2}$O 3$_{12} - 3_{03}$ &3160 \\
            &             &             & 1342243624 & SPIRE FTS&2530 \\
            &             &             & 1342207776 & PACS Spec-red &2263 \\
            &             &             & 1342207777 & PACS Spec-blue&1135 \\
OH~30.1-0.7  & 18 48 41.91 & -02 50 28.3 & 1342218431 & CO 5-4   &118 \\
            &             &             & 1342229923 & CO 9-8   &921\\
\hline
\end{tabular}
\end{table*}

We also have {\it Herschel} SPIRE FTS \citep{griffin10} observations 
(calibration file SPG v9.1.0) of WX Psc and OH~26.5+0.6 (P.I. M. Barlow)
as well as archival PACS \citep{poglitsch10} spectra
of these objects (calibration file SPG v6.1.0) from the MESS
guaranteed time program \citep{groen11}, 
which are used to constrain the 
spectral energy distribution (SED) of individual objects.
The PACS spectra were extracted from the central spaxel with no
further correction hence the absolute flux for long wavelengths may be
underestimated.
The SPIRE data for OH~26.5+0.6 show a strong flux excess at the high frequency 
end of the SLW subspectra. This effect is due to the pick-up of extended 
emission, either
from the source itself, or from its background, by the multi-moded beam
applying to this range of the SLW array \citep[see e.g.,][]{swinyard10}.
Inspection of the spectra in the surrounding pixels confirms the compactness
of the CO emission detected in the central pixel. The CO envelope is also
known to be compact from the BIMA map \citep{fong02}. We
therefore  corrected for this excess using the smoothed average of the
spectra collected over the first ring of surrounding detectors. This correction
yields the expected continuity
between the calibrated spectra of both SLW and SSW central pixels.
The spectral resolution of both the FTS and PACS data makes it impossible 
to study 
individual CO line profiles and the line fluxes can be affected by 
blending from other molecules, notably, H$_{2}$O. The CO line fluxes are
calculated by fitting a Gaussian to individual lines in the PACS spectra 
and sinc functions to unapodized SPIRE spectra (Table~\ref{tab_linefluxes}). 
The error estimates of these
line fluxes are due to the noise of the baseline and do not include
possible blends with other molecules. It can be seen that for high-J lines 
in the PACS range, the baseline gets increasingly noisy.
To better constrain the mass 
loss rate throughout the envelope, we extracted archival CO ground-based data 
from APEX, JCMT and IRAM \citep{debeck10}.

\begin{table}[h]
\caption{CO line fluxes from Guassian fitting the SPIRE (up to J=13-12)
and PACS observations.}
\label{tab_linefluxes}
\begin{tabular}{lll}
\hline \hline
Transition & \multicolumn{2}{c}{Flux (10$^{-18}$ W m$^{-2}$)} \\
\hline
   & WX Psc & OH~26.5+0.6  \\
\hline
5-4   & 38.3$\pm$4.9  &  32.3$\pm$4.5 \\
6-5   & 37.2$\pm$4.9  &  17.6$\pm$4.5 \\
7-6   & 45.3$\pm$4.9  & 19.5$\pm$4.6 \\
8-7   & 52.8$\pm$4.9  & 32.3$\pm$4.5 \\
9-8   & 60.7$\pm$9.3  & 28.1$\pm$11.1 \\
10-9  &169.4$\pm$9.2  & 117.1$\pm$11.1 \\
11-10 & 73.5$\pm$9.2  &  23.3$\pm$11.1 \\
12-11 & 75.8$\pm$9.2  &  40.8$\pm$11.1 \\
13-12 & 86.7$\pm$9.3  &  39.8$\pm$11.1 \\
18-17 & 67.0$\pm$8.1   & 52.9$\pm$10.0 \\
19-18 & 50.8$\pm$2.2   & 21.6$\pm$10.0 \\
20-19 & 54.0$\pm$5.4   & 38.9$\pm$10.0 \\
21-20 & 54.0$\pm$4.3   & 30.2$\pm$30.2 \\
22-21 & 73.4$\pm$11.8  & 18.4$\pm$18.4 \\
24-23 & 103.7$\pm$17.3 & - \\
25-24 & 130.0$\pm$17.3 & 32.4$\pm$18.4 \\
26-25 & 75.6$\pm$21.0  &  - \\
27-26 & 89.6$\pm$4.9   &  - \\
29-28 & 129.6$\pm$27.0 &  - \\
30-29 & 81.0$\pm$60.0  &  - \\
31-30 & 86.4$\pm$60.0  &  - \\
32-31 & 70.2$\pm$60.0  &  - \\
\hline
\end{tabular}
\end{table}

\section{Dust and gas mass loss rates}

In order to derive the dust mass loss rate for each object, we fit the
SED using the archival data from the Infrared
Space Observatory (ISO) and {\it Herschel} as well as published photometry. 
Since the main dust species is silicates, which are responsible
for driving the wind, we derived the dust mass loss rate, 
$\dot{M}_{d}$, by mainly fitting the
silicate 10 and 20~$\mu$m features and use the photometric points as a guide
for the overall shape (Fig~\ref{fig-sed}). 
The radiative transfer code used is based on 
\cite{haisch79} and the dust opacity is from \cite{justtanont06}.
The input parameters for the SED fitting (Table~\ref{tab2}) include 
the stellar luminosity
as calculated from the stellar effective temperature (T$_{*}$) and radius
(R$_{*}$). The expansion velocity, v$_{e}$, is that measured 
from CO observations. The distance, D, is taken from the literature. 
Assuming spherical symmetry and that the dust and
gas are fully momentum coupled, we can derive the dynamical mass loss rate,
$\dot{M}_{dyn}$,
due to the dust driven wind \citep[see][]{goldreich76,justtanont06}. One of the
unknown parameters in this calculation is the mass of the star, which we
assume to be 5~M$_{\odot}$ in all cases. This stems from the fact that
these stars show the signature of hot-bottom burning via their low 
$^{12}$C/$^{13}$C ratio \citep{delfosse97}. 
This process operates in stars more massive than
$\sim$ 5~M$_{\odot}$\citep{boothroyd93}. 
We discuss this point later in section 4.
The derived dynamical mass loss rate
is the total gas mass loss rate which is driven by the
dust grains to the observed terminal velocity.
It can be seen that the derived dynamical mass loss rates for our sample
of OH/IR stars are relatively high.

\begin{table}[h]
\caption{Input parameters for SED and dust-drag modelling. The units
of the dust mass loss rate ($\dot{M}_{d}$) and the
derived dynamical mass loss rate ($\dot{M}_{dyn}$) are 
M$_{\odot}$ yr$^{-1}$.}
\label{tab2}
\begin{tabular}{lccccc}
\hline \hline
             & WX Psc & OH127 & AFGL5379 & OH26 & OH30 \\
\hline
T$_{*}$ (K)  & 2200   & 2300   & 2200   & 2200   & 2200 \\
R$_{*}$ (cm) & 4.8E13 & 9.0E13 & 5.0E13 & 6.0E13 & 5.0E13 \\
v$_{e}$ (km s$^{-1}$) &
               19.8   & 12.7   & 18.3   & 15.0   & 18.1  \\
D (kpc)      & 0.68   & 2.8    & 0.58   & 1.37   & 1.75   \\
$\dot{M}_{d}$ &
               1.8E-7 & 2.0E-6 & 1.6E-6 & 2.0E-6 & 1.8E-6 \\
$\dot{M}_{d}/\dot{M}_{g}$ &
               9.4E-3 & 2.2E-3 & 1.0E-2 & 6.2E-3 & 1.0E-2 \\
$\dot{M}_{dyn}$ &
               1.9E-5 & 9.2E-4 & 1.6E-4 & 3.2E-4 & 1.8E-4 \\
\hline
\end{tabular}
\end{table}

A correlation between observed CO line intensities and mass loss rates
has been well established \citep[e.g.,][] {knapp85, olofsson93, debeck10}. 
In spite of this, 
a number of extreme OH/IR stars which exhibit silicate absorption bands 
observed by \citet{heske90} showed anomalously weak CO J=1-0 
and 2-1 emission. These authors proposed scenarios where the high 
density is efficient
in cooling the gas or that the star had increased its
mass loss rate recently and the high density has not propagated out
to where the lowest rotational transitions originate.
In order to calculate the CO emission, we use a code based on 
the work by \cite{schonberg86}. For the mass loss rate, We used the values
derived from the SED fitting and dynamical calculation as an input
for individual object.
For very optically thick envelopes,
a change in the gas density has much less effect than 
a change of gas temperature of the same factor, 
hence in calculating the CO emission in these
stars, the input gas temperature affects the relative distribution
of population and hence the calculated line fluxes.
The gas temperature is calculated from the
energy balance between the dust-drag heating and the cooling by
adiabatic expansion and molecular cooling, following \cite{goldreich76}.
We also include the effect of photoelectric heating due to
photoejected electrons in the outer part of the envelope \citep{tielens85}.
The main uncertainty is due to the H$_{2}$O cooling rate which is crudely
estimated. 
%
%
The difficulty with the full radiative transfer calculation for H$_{2}$O
is due to numerous transitions within the temperature range of the 
circumstellar environment (T $\sim$ 10-2000~K) and that theses lines
have very high optical depths, so that our current code fails to converge.
However, progress is being made in modelling H$_{2}$O in S-stars
\citep[][]{schoier11,danilovich13} and in O-rich stars 
\citep{maercker09} with low 
mass loss rates. The H$_{2}$O cooling in OH/IR stars will be addressed in a
future work, together with the HIFI observations.
For our calculations, the H$_{2}$O cooling rate is unconstrained and we 
vary the gas temperature until it gives 
reasonable fits to all
the rotational CO lines observed. 
Hence, we can constrain the
gas kinetic temperature of the high-density superwind rather well.
In our sample of objects observed
with {\it Herschel}, the CO emission indicates that the mass loss rate
in all but one star (WX Psc) has recently increased. 
Using the calculated dynamical mass loss rate for each system, we
derived the $^{12}$CO abundance relative to H$_{2}$ which is relatively 
constant for all the sources, i.e., all have a value of 3\,10$^{-4}$.

\subsection{WX Psc (IRC+10011)}

This star shows a moderate dust mass loss rate compared to the others in our
sample as the 10$\mu$m silicate feature shows self-absorbed emission. 
The derived
dynamical mass loss rate is 1.7\,10$^{-5}$ M$_{\odot}$ yr$^{-1}$. The SED is
best fitted by the dust grains which have an emissivity dependence 
$\lambda^{-1.5}$
(see Fig.~\ref{fig-sed}). The model fit to individual CO transitions from 
ground-based telescopes and {\it Herschel} assumes an outflow with
a constant mass loss rate (Fig.~\ref{co_irc10011}). We note that 
\cite{decin07} reported a variation in mass loss rate by fitting the
ground-based CO observations.
Their reported JCMT observations of this object are not consistent with 
the HIFI J=6-5 line and the JCMT J=7-6 line is weaker than that observed with 
the smaller APEX telescope \citep{debeck10}.
This suggests that there is a problem with calibration for JCMT observations
of this star. The object has been observed to be spherically symmetric on
a large scale \citep{mauron06} but with a small scale extension
\citep[e.g.,][]{hofmann01,inomata07,ladjal10}.
In view of this, we take the simplest approach of
fitting the HIFI and ground-based lines with a constant mass loss rate. 

\begin{figure}[h]
\centering
\resizebox{\hsize}{!}{\includegraphics{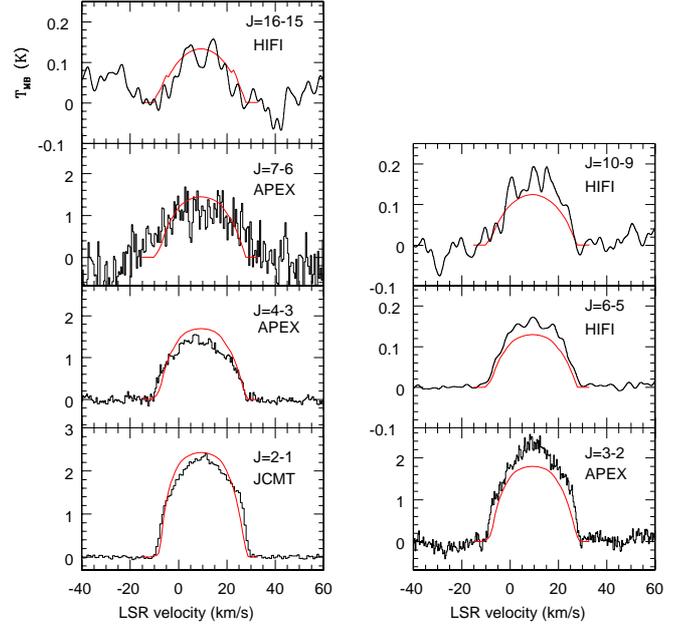}}
\caption{Observed $^{12}$CO rotational spectra (black histogram) of WX Psc with
the model fit (smooth red line).
}
\label{co_irc10011}
\end{figure}

PACS spectrometer and SPIRE FTS data were extracted for this object. 
A series of CO  
rotational lines can be seen. The measured line fluxes are plotted in 
Fig.~\ref{spire_flux} as crosses and open symbols, respectively. 
After careful baseline subtraction, it can be seen that the line fluxes of
CO 6-5 as measured by HIFI and SPIRE agree extremely well.
The high observed J=10-9 flux from SPIRE 
is due to a blend with the H$_{2}$O
3$_{12}$-2$_{21}$ transition which is about twice as strong as the CO line 
\citep{justtanont12}. Our model of the CO emission follows the general rising
trend of the fluxes from the {\it Herschel} observations.
The model, however, predicts lower line fluxes than observed with SPIRE,
while the observed PACS CO lines become increasingly noisy.
The mismatch between the SPIRE observation and the model may be due to
the assumption of a constant mass loss rate and the assumption of 
spherical symmetry in the calculation.

\begin{figure}[h]
\centering
\resizebox{\hsize}{!}{\includegraphics{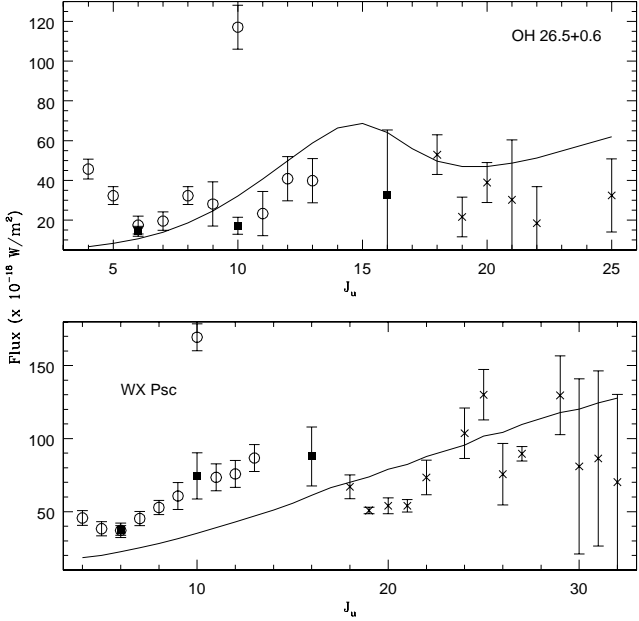}}
\caption{The CO line fluxes taken SPIRE (open circles), HIFI (filled
squares) and PACS (crosses) are shown together with the predicted fluxes 
from the model
which is the best fit to both ground-based observations and HIFI for
WX Psc (bottom) and OH~26.5+0.6 (top). The high SPIRE flux of J=10-9
is due to a blend with a strong H$_{2}$O line.
}
\label{spire_flux}
\end{figure}

HIFISTARS observed four transitions of $^{13}$CO J=6-5, 9-8, 10-9 and
16-15 \citep{justtanont12}, although the highest transition is extremely noisy
and can be viewed as an upper limit (Fig.~\ref{13co_irc}). 
Using the same parameters as for the $^{12}$CO model, we
derive an abundance for $^{13}$CO relative to H$_{2}$ of (3$\pm$1)\,10$^{-5}$,
i.e., $^{12}$C/$^{13}$C  $\sim$ 10$\pm$4 (Table~\ref{13co}). The
estimated uncertainty of this ratio is a combination of the uncertainty 
of the derived $^{13}$CO/H$_{2}$ ratio and a canonical 20\% uncertainty for 
$^{12}$CO abundance.
This ratio is at the lower end of the estimated range of 10-35 for a sample of 
O-rich AGB stars \citep{milam09}.
\begin{figure}[h]
\centering
\resizebox{\hsize}{!}{\includegraphics{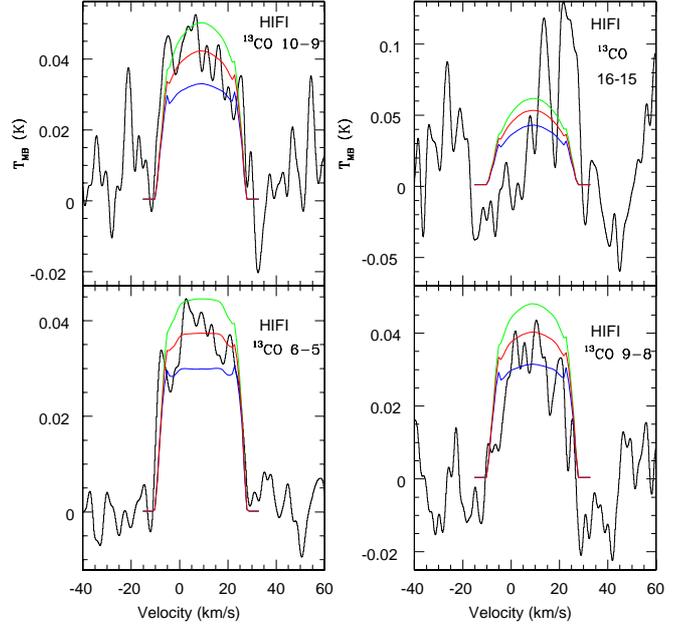}}
\caption{Observed HIFI $^{13}$CO rotational spectra (black histogram) of 
WX Psc with the model fit (smooth red line). Also plotted are the 
abundance limits for $^{13}$CO/H$_{2}$ of 2\,10$^{-5}$ (blue line)
and 4\,10$^{-5}$ (green line).
}
\label{13co_irc}
\end{figure}

\subsection{OH~127.8+0.0}

The infrared spectrum of this object displays deep silicate absorption
bands at 10 and 20~$\mu$m, indicating a very high current mass loss rate
(Fig.~\ref{fig-sed}). The derived dynamical mass loss rate of the star is 
9.2\,10$^{-4}$ M$_{\odot}$ yr$^{-1}$.

However, using a constant mass loss rate results in overestimating the
ground based CO observations. In order to fit both the {\it Herschel} HIFI
and IRAM \citep{delfosse97} observations, the mass loss rate in the outer 
part where the J=1-0 and 2-1 originate is taken to be a fraction of the
derived dynamical mass loss rate (Fig~\ref {co_oh127}).
The radius where the superwind ends, r$_{sw}$ (Table~\ref{13co}),
is 130 R$_{*}$, i.e., 1.2\,10$^{16}$ cm and the mass loss rate beyond this 
radius is set to 2.8\,10$^{-6}$ M$_{\odot}$ yr$^{-1}$. 
The HIFI lines taken from
the SUCCESS program are noisy due to shorter integration time compared
with spectra taken from HIFISTARS. The fits
to these lines, although not perfect, are reasonable given the relatively
low signal-to-noise ratio. Recently, \cite{lombaert13} studied the mass loss
of this star and concluded that in order to fit all the available
CO data, the mass loss rate in the outer part must be lower compared to the
inner superwind.

\begin{figure}[h]
\centering
\resizebox{\hsize}{!}{\includegraphics{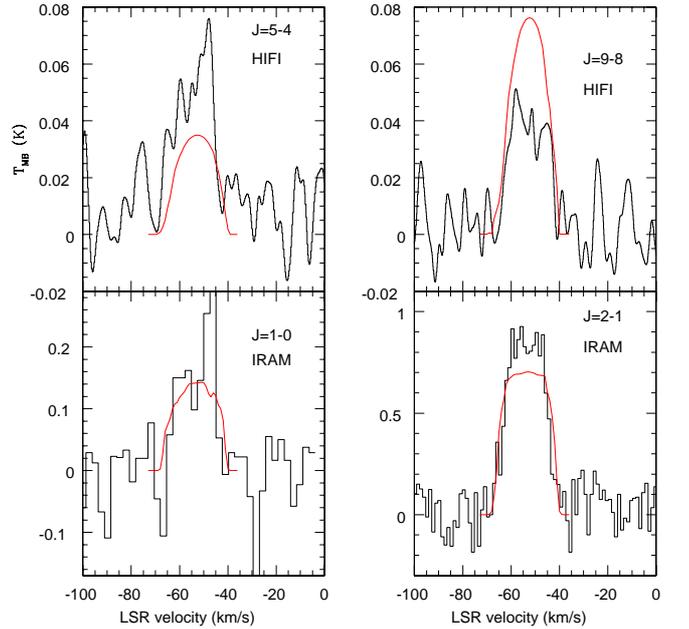}}
\caption{Observed $^{12}$CO rotational spectra (black histogram) of 
OH~127.8+0.0 with the model fit (smooth red line) which has an inner superwind
surrounded by a tenuous outer wind.
}
\label{co_oh127}
\end{figure}

\citet{delfosse97} also observed $^{13}$CO 1-0 and 2-1 from this object. 
Using the same approach as for WX Psc, we used the same input parameters 
except for the abundance to fit the two lines (Fig.~\ref{13co_oh127_30}). 
We derive a $^{13}$CO/H$_{2}$ abundance of 1.5$\pm$0.5\,10$^{-4}$, i.e.,
a $^{12}$C/$^{13}$C ratio of 2$\pm$1 (Table~\ref{13co}).
This is consistent with the central star having undergone hot-bottom 
burning, destroying $^{12}$C and reaching the equilibrium
$^{12}$C/$^{13}$C ratio of $\sim$ 3. \citep{lattanzio03}.

\begin{figure}[h]
\centering
\resizebox{\hsize}{!}{\includegraphics{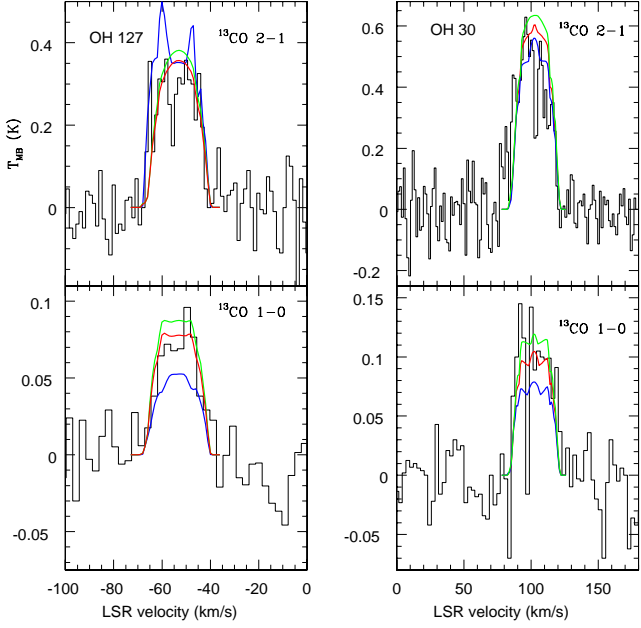}}
\caption{Observed $^{13}$CO spectra from IRAM (black histogram)  
for OH~127.8+0.0 (left panels) and OH~30.1-0.7 (right panels)
with the model fit (smooth red line) and the upper (green)
and lower (blue) limits to the abundances for the species (see text). 
}
\label{13co_oh127_30}
\end{figure}

\subsection{AFGL~5379}

For this object we use a distance of 580~pc, taken from \citet{veen89},
rather than the larger distance of 1190~pc  derived by \citet{yuasa99}. 
As can be seen from the SED in Fig.~\ref{fig-sed}, its flux 
level is an order of magnitude higher than the other extreme OH/IR stars
in the sample and is similar to that of WX Psc. The infrared spectrum shows
deep silicate absorption features, reflected in the high derived dynamical mass
loss rate of 1.6\,10$^{-4}$ M$_{\odot}$ yr$^{-1}$.
The far-IR silicate dust efficiency is best fitted with a slope of 
$\lambda^{-1.5}$, as in the case of WX Psc.

\begin{figure}[h]
\centering
\resizebox{\hsize}{!}{\includegraphics{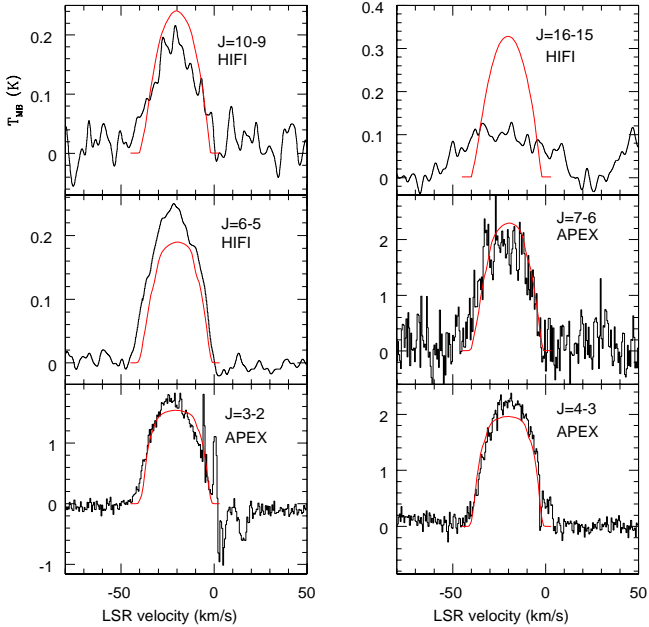}}
\caption{Observed $^{12}$CO rotational spectra (black histogram) of AFGL~5379
with the model fit (smooth red line) which has an inner superwind
surrounded by a tenuous outer wind.
}
\label{co_gl5379}
\end{figure}

In order to fit the observed CO lines, the superwind has to be truncated at
a radius of 10$^{16}$ cm. The outer mass loss rate for this model is 
3.9\,10$^{-6}$ M$_{\odot}$ yr$^{-1}$. The model gives reasonable fits to
all the lines (Fig.~\ref{co_gl5379}), 
except for the J=16-15 line where the observation is highly
affected by the sinusoidal effect seen in HIFI observation in bands 6 and 7
(1.44-1.92~THz) due to standing waves in the instrument 
\citep{roelfsema12}. However, further careful analysis confirms the 
line is much weaker than the model. The misfit of the model indicates that
the simple assumption of spherical symmetry may not apply close to the star.

\begin{figure}[h]
\centering
\resizebox{\hsize}{!}{\includegraphics{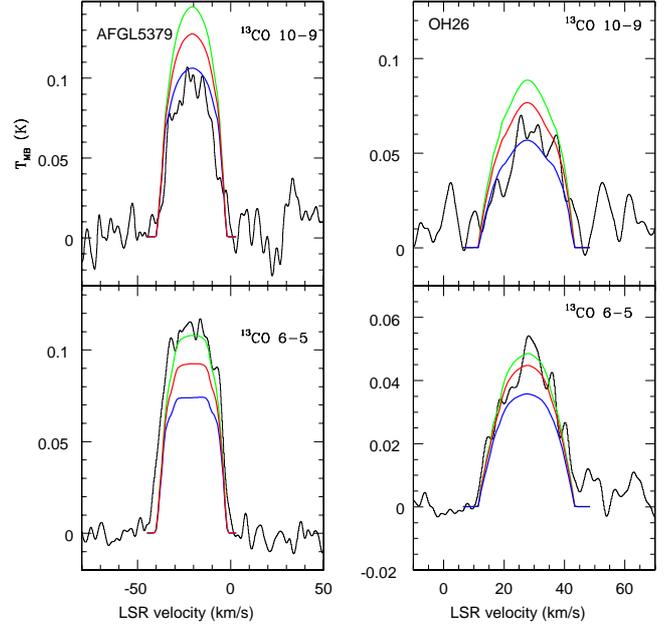}}
\caption{Observed HIFI $^{13}$CO rotational spectra (black histogram) with 
the model fit (smooth red line) and with the upper (green line) and lower
(blue line) limits to the $^{12}$C/$^{13}$C abundance ratio 
for AFGL~5379 (left panels) and OH~26.5+0.6
(right panels).
}
\label{13co_gl_oh26}
\end{figure}

The fit to the HIFI observations of the two $^{13}$CO transitions leads to a
$^{13}$CO/H$_{2}$ 
abundance of (1.1$\pm$0.4)\,10$^{-5}$ (Fig~\ref{13co_gl_oh26}). 
Both lines cannot be well fitted with a single abundance, however.
The resulting ratio of $^{12}$C/$^{13}$C is
of 27$\pm$11. This high value could mean that the star has either 
just started the process of hot-bottom burning and the $^{13}$C
has not been significantly produced or the process has already
taken place and the central star has entered a phase where 
$^{12}$C is being synthesized again.
Further evidence that
hot-bottom burning process has started in this star is the non-detection
of the H$_{2}^{18}$O line (see section 4).

\subsection{OH~26.5+0.6}

Extensive studies have been done of OH~26.5+0.6
over the years in order to establish the
superwind radius (r$_{\rm sw}$, i.e., 
the radius where the intense mass loss rate has propagated outwards).
\citet{fong02} resolved the circumstellar envelope in
the CO J=1-0 line, giving an outer radius of 7\,10$^{16}$ cm but did not
resolve the superwind. Mid-infrared interferometric observations established
the full-width-half maximum of the dusty envelope of $\sim$ 280 milliarcsec 
\citep{chesneau05}.
At a distance of 1.37 kpc \citep{langevelde90}, 
this translates to 6\,10$^{15}$ cm. \citet{driebe05}
observed this object in the K$^\prime$ band and resolved the inner radius
of the dust shell to be between 29.3~mas at the minimum phase and 69.5~mas
during the maximum phase.

\begin{figure}[h]
\centering
\resizebox{\hsize}{!}{\includegraphics{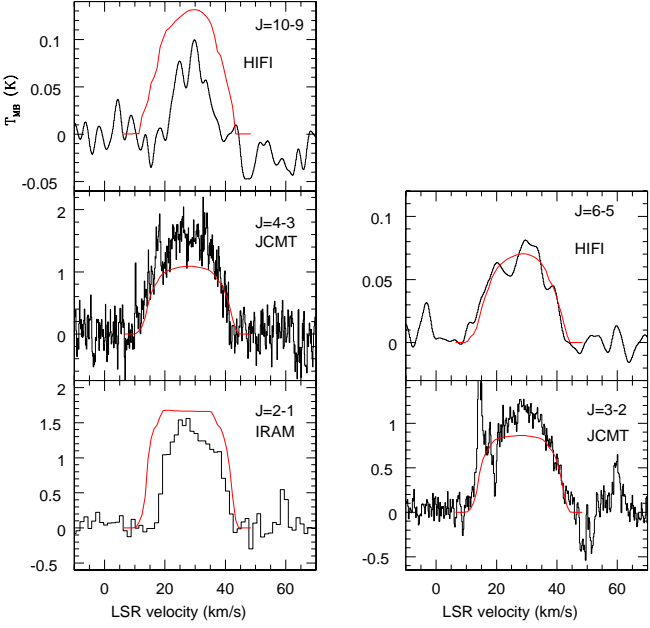}}
\caption{Observed $^{12}$CO rotational spectra (black histogram) of OH~26.5+0.6 
with the model fit (smooth red line) with an inner superwind
surrounded by a tenuous outer wind.
}
\label{co_oh26}
\end{figure}

From fitting the SED and solving the equation of motion between the dust and
gas interaction, we derive a dynamical mass loss rate of 3.2\,10$^{-4}$
M$_{\odot}$ yr$^{-1}$. The continuum fluxes observed by {\it Herschel}
using PACS and SPIRE (Fig.~\ref{fig-sed})
follow the ISO fluxes reasonably well, except for
the longest wavelength channel of SPIRE (Fig~\ref{fig-sed}).
To fit the observed CO lines, we limit the superwind radius to 9\,10$^{15}$
cm, with a mass loss rate beyond this radius of 1.6\,10$^{-6}$
M$_{\odot}$ yr$^{-1}$ (Fig~\ref{co_oh26}). 
The IRAM J=2-1 line has the interstellar line component removed \citep{heske90}
hence its blue wing is missing. The HIFI spectrum of CO J=10-9
displays a much narrower line than expected. It is not clear why this is the
case. 
It is possible that the wind acceleration is not as steep,
but rather that in the region where the CO 16-15 originates the velocity 
is lower than the result of the dust-driven wind would suggest.
In Fig.~\ref{spire_flux}, it can be seen that the CO 10-9 line from SPIRE
show a much higher flux than the corresponding HIFI line due to the 
strong H$_{2}$O blend.  
The high-J CO lines observed by PACS are noisy due to the low flux.  

From HIFISTARS, two $^{13}$CO transitions were detected and our derived 
abundance for $^{13}$CO/H$_{2}$ for this object is (1$\pm$0.5)\,10$^{-5}$. 
The uncertainty of this abundance is relatively high since the 6-5 line
is more optically thick compared to the 10-9 transition.
This results in a $^{12}$C/$^{13}$C ratio of 30$\pm$16 which is relatively high
for our sample but with a large uncertainty.

\subsection{OH~30.1-0.7}

The star has been observed by ISO to have deep silicate absorption bands, like
most other stars in our sample. It is also be seen to have water-ice and
crystalline silicates \citep{justtanont06}. 
Our derived dynamical mass loss rate is 1.8\,10$^{-4}$ M$_{\odot}$ yr$^{-1}$.

\begin{figure}[h]
\centering
\resizebox{\hsize}{!}{\includegraphics{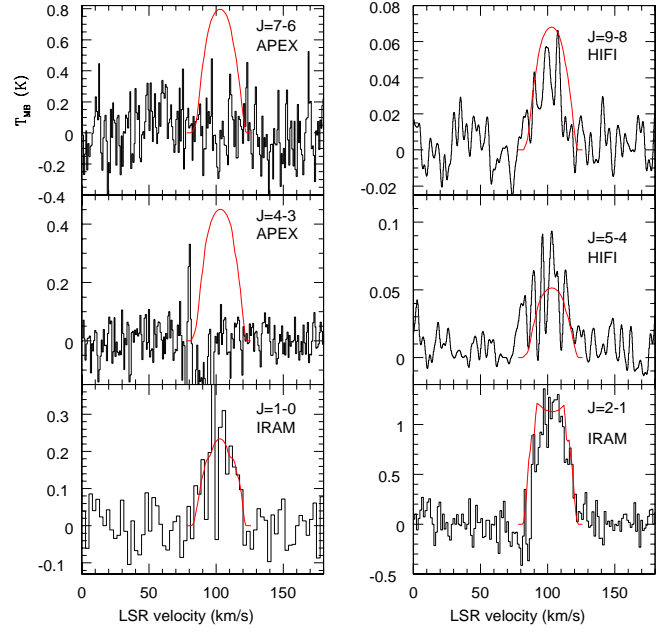}}
\caption{Observed $^{12}$CO rotational spectra (black histogram) of OH~30.1-0.7
with the model fit (smooth red line) with an inner superwind
surrounded by a tenuous outer wind.
}
\label{co_oh30}
\end{figure}

The ground-based CO observations taken at APEX reported by \citet{debeck10}
showed a non-detection (Fig~\ref{co_oh30}) which is most likely due to the
mispointing of the telescope (15$^{\prime\prime}$ off in the RA direction
from checking the header information of the APEX observations)
as the source was readily detected by
IRAM \citep{delfosse97} and {\it Herschel}.
To fit the IRAM and HIFI observations, the radius
of the superwind is limited to 2.5\,10$^{16}$ cm with the mass loss rate
in the outer region falling to 1.8\,10$^{-6}$ M$_{\odot}$ yr$^{-1}$
(Fig.~\ref{co_oh30}). The derived abundance of $^{13}$CO/H$_{2}$ is
(8$\pm$0.3)\,10$^{-5}$, resulting in a $^{12}$C/$^{13}$C 
ratio from fitting the $^{13}$CO IRAM observations 
(Fig.~\ref{13co_oh127_30})
for this star of 4$\pm$1, which is close to the expected 
equilibrium value for stars which have experienced hot-bottom burning
\citep{boothroyd93}.

\begin{table}[h]
\caption{Derived superwind radius, r$_{\rm sw}$ and $^{13}$CO abundances for 
the stars in this study. The derived $^{12}$CO abundance is 3\,10$^{-4}$.}
\label{13co}
\begin{tabular}{lccc}
\hline \hline
  & r$_{\rm sw}$ (cm) &  [$^{13}$CO]/[H$_{2}$] & $^{12}$C/$^{13}$C \\
\hline
WX Psc       & -      & (3.0$\pm$1.0)E-5 & 10$\pm$4 \\
OH~127.8+0.0 & 1.2E16 & (1.5$\pm$0.5)E-5 & 2$\pm$1 \\
AFGL~5379    & 1.0E16 & (1.1$\pm$0.4)E-5 & 27$\pm$11 \\
OH~26.5+0.6  & 9.0E15 & (1.0$\pm$0.5)E-5 & 30$\pm$16 \\
OH~30.1-0.7  & 2.5E16 & (8.0$\pm$0.4)E-5 & 4$\pm$1 \\
\hline
\end{tabular}
\end{table}

\section{Discussion}

From observations and CO modelling, we derived mass loss rates
and CO abundances relative to H$_{2}$, as well as those for $^{13}$CO 
for all stars in our sample. Here, we discuss the implications of our 
finding.

\subsection{The relative abundances C and O isotopes}

The solar value of $^{12}$C/$^{13}$C is estimated to be 87 \citep{scott06},
which is similar to the value derived from meteorites \citep{anders89}.
It has been reported that the galactic
ratio appears to be dependent on the distance from the galactic centre
\cite[see e.g.,][]{henkel94,wilson94}.
The interstellar ratios estimated within 100 pc along different
lines of sight vary between 62 and 98 \citep{casassus05}. 
A solar isotopic ratio of 480 for $^{16}$O/$^{18}$O has been determined by 
\cite{scott06} 
while the solar $^{16}$O/$^{17}$O ratio is taken to be 2600
with a large uncertainty due to the weakness of the C$^{17}$O lines
\citep{asplund09}.

There are extensive observations of CO and CN
in C-rich AGB stars which show that the $^{12}$C/$^{13}$C ratio has a range
between 30 and 150 \citep[e.g.,][]{lambert86,ohnaka96,schoier00} with a few
stars showing a low $^{12}$C/$^{13}$C ratio. 
\cite{ohnaka99} found small ratios 
for J-type C-stars, between 2 and 10.
The values derived for circumstellar envelopes of
O-rich AGB stars are relatively limited and have a range of 10 to 80
\citep{bujarrabal94,milam09}. 
For sufficiantly high-mass AGB stars, $^{13}$C can be produced by CNO-cycle
hot-bottom burning of the third dredge-up $^{12}$C and then transported 
to the photosphere to become part the the circumstellar envelope 
due to mass loss
processes during the AGB phase. The ratio of the carbon isotopes can
be used as a tracer of stellar mass as $^{13}$C is readily 
synthesized in high mass AGB stars \citep{lattanzio03}.

\begin{figure}[h]
\centering
\resizebox{\hsize}{!}{\includegraphics{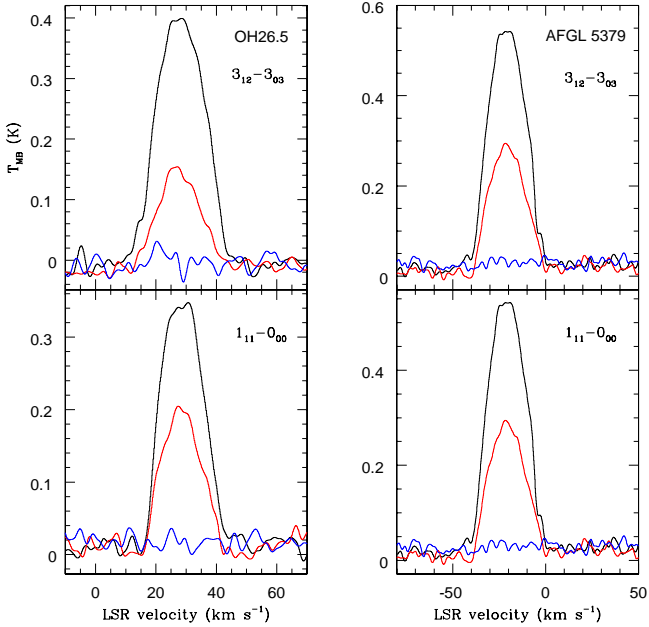}}
\caption{HIFISTARS spectra of OH~26.5+0.6 (left) and AFGL~5379 (right) showing 
the transitions 3$_{12} - 3_{03}$ and 1$_{11} - 0_{00}$
of H$_{2}^{16}$O (black), H$_{2}^{17}$O (red) and
H$_{2}^{18}$O (blue). The H$_{2}^{18}$O lines are not detected in either
source in both ortho- and para-H$_{2}$O.
}
\label{h2o}
\end{figure}

Three of the five stars in our sample 
have low $^{12}$C/$^{13}$C ratios (Table~\ref{13co})
as derived from the CO line fluxes, based on both ground based 
and HIFI observations. 
Due to the high mass loss rates, even $^{13}$CO lines 
can be optically thick.
It is curious to note that the ratios
derived from the IRAM observations are lower than those derived
from the HIFI observations. However, the derived abundance ratios from
high-J lines have larger uncertainties associated with them due to
the high $^{13}$CO line opacity in the superwind.
A low $^{12}$C/$^{13}$C ratio is expected
when a star undergoes hot-bottom burning.
The lower mass limit for this process to occur is $\textgreater$ 4~M$_{\odot}$
\citep{sackmann92}. Theoretical calculations of the third dredge-up
for stars more massive than 5~M$_{\odot}$ show that this ratio initially
increases from $\sim$ 17 to 30 and then drops down to a value of 3 once
the temperature at the base of the convective envelope is hot enough
to produce significant $^{13}$C from $^{12}$C \citep{lattanzio03}. 
This value increases again after the end of this phase.
The implication of our derived $^{12}$C/$^{13}$C ratio is
that these extreme OH/IR stars have originated from the high mass end of 
intermediate-mass stars.

Although $^{12}$C/$^{13}$C ratios have been studied in AGB stars,
very little work has been done on the isotopic ratios of $^{18}$O and
even less on $^{17}$O due to thier low abundances. Thanks to the
high sensitivity of {\it Herschel}, we can begin such studies 
using H$_{2}$O molecules.
As part of the HIFISTARS observations, three of the stars in our sample have 
been observed with a frequency setting which included the 
1$_{11} - 0_{00}$ (para-H$_{2}$O) and 3$_{12} - 3_{03}$ (ortho-H$_{2}$O)
transitions for H$_{2}^{16}$O, H$_{2}^{17}$O and H$_{2}^{18}$O 
(Table~\ref{tab-sources}).
\cite{justtanont12} reported the HIFI spectra of WX Psc which 
showed line flux ratios for  H$_{2}^{18}$O/H$_{2}^{17}$O of 1.5 for both
ortho and para lines.
For AFGL~5379 and OH~26.5+0.6, it
can be clearly seen here that the H$_{2}^{16}$O and H$_{2}^{17}$O are
detected while the H$_{2}^{18}$O line is not detected above the noise
(Fig.~\ref{h2o}). 
The estimated line flux ratios of H$_{2}^{18}$O/H$_{2}^{17}$O
for AFGL~5379 are 0.05 and 0.09  and for OH~26.5+0.6 are 0.13 and 0.09 
for the ortho and para lines, respectively. These values are well below
the solar $^{18}$O/$^{17}$O ratio of $\sim$ 5 and the interstellar
value of $\sim$ 3.5 \cite{wilson94}.
The non-detection of H$_{2}^{18}$O in AFGL~5379 and OH~26.5+0.6 and the
almost equally strong H$_{2}^{17}$O and H$_{2}^{18}$O line fluxes in WX Psc
provide evidence that H$_{2}^{18}$O is
destroyed in the hot-bottom burning process which operates
in more massive AGB stars \citep{lattanzio96}. 
This result gives a strong indication of the
lower limit to the stellar mass which is not as model dependent as for 
the case of the derived $^{12}$C/$^{13}$C ratios.

\subsection{The superwind}

The phenomenon of a superwind has been postulated because a constant
mass loss rate does not remove enough stellar mass during the AGB lifetime
to allow low- and intermediate-mass stars to evolve into planetary nebulae.
\cite{vassiliadis93} showed that a superwind develops during the last thermal
pulses while the stellar luminosity and pulsation period increase. For
a 5-M$_{\odot}$ star, the superwind can last 10$^{5}$ yr, i.e.
half of the thermal-pulse AGB lifetime.
From our modelling, the radius of this superwind, r$_{\rm sw}$,
(Table~\ref{13co}) was derived 
for each object from fitting high- and low-J CO lines. Assuming a constant
expansion velocity for these lines, we can calculate the lifetime of the
superwind for individual stars. This implies that the
stars in our sample have entered the superwind phase
only in the past few hundred years, i.e., 0.3\% of the theoretical lifetime
of the superwind phase. 
During the Infrared Astronomical Satellite ({\it IRAS}) all sky survey, 
50 000 low resolution spectrometer (LRS) spectra were taken, of which 
about 150 show silicate absorption with a blue continuum (class 3n) which are
associated with OH/IR stars \citep{likkel89}. However, there may still be some
contamination in this classification. \cite{suh09} 
searched the {\it IRAS} LRS catalogue
and found that out of 1400 spectra of O-rich stars and 931 C-stars, 180
objects showed the 10~$\mu$m silicate feature in absorption, a similar
number to that found by \cite{likkel89}. 
This indicates that from an unbiased all-sky
survey the fraction of extreme OH/IR stars is $\sim$ 8\% of
the AGB population -- 
much higher than the ratio of the estimated superwind lifetime to the total
thermal-pulse time on the AGB.
When stars enter a superwind phase, they become very bright
in the infrared and luminous in OH masers -- OH~26.5+0.6 and AFGL~5379
are among the AGB stars which exhibit the strongest OH 1612 MHz masers
\citep{telintel89}.
Even though these objects are confined to the galactic plane and hence 
suffer from heavy extinction as well as from confusion due to the
high density of stars in the plane, the luminosity bias means that
they are easily picked up in maser and infrared surveys. This could partially 
be the reason why we detect an enhanced fraction of these objects.

\section{Summary}

From fitting the SEDs of a sample of extreme OH/IR stars, we have derived 
very high mass loss rates, reflecting the absorption seen in the 10$\mu$m
silicate dust feature. Using CO as a mass loss tracer, from ground based
observations and those obtained by {\it Herschel}, we propose
that four out of five stars in our sample have recently increased their 
mass loss
rates. The high-J CO lines probe the warm envelope inside the superwind
while the low-J lines originate from outer cooler regions where
the mass loss is lower. The timescale since the start of the superwind phase
is estimated to 
vary between $\sim$ 200-500 years. From the ground based observations done
so far, there is no evidence that there has been a superwind phase earlier 
than observed by us, i.e., there is no obvious evidence of multiple 
shells of mass loss around the objects \citep{cox12}.

The derived $^{12}$C/$^{13}$C ratios in some objects 
are much lower than the canonical value
of $\sim$ 20-30 observed for stars with initial masses of 4-6 M$_{\odot}$
after the third dredge-up \citep{lattanzio03}. Due to the effect of 
hot-bottom burning, this ratio can be reduced to closer to 3-4 over a timescale
of $\sim 10^{5}$ yr.
Our modelling gives a range for this ratio of between 2-30, with some
stars having a large uncertainty on this ratio.
The low observed $^{18}$O/$^{17}$O ($\sim$ 0.1) ratios for the stars 
in our sample compared, with the solar value of $\sim$ 5,
give a strong indication that these stars 
have undergone the process of hot-bottom burning, implying
that they are massive ($\geq$ 5 M$_{\odot}$) AGB stars.

These stars are seen to be presently losing mass at prodigious rates
of $\sim 10^{-4}$ M$_{\odot}$ yr$^{-1}$.
High mass loss rate AGB stars may be the dominant source of dust and gas 
return to the ISM \citep{matsuura09} and these 
stars can have a significant impact on the chemical and dust enrichment 
of the interstellar medium.
They are thus crucial in studies of galactic chemical evolution.

\begin{acknowledgements}
This research is party funded by the Swedish National Space Board. We would
like to thank Leen Decin for her valuable comments to the manuscript.
We also thank both the referee and the editor (Malcolm Walmsley) for further
comments for improvement of this paper.

HIFI has been designed and built by a consortium of institutes and university 
departments from across Europe, Canada and the United States under the 
leadership of SRON Netherlands Institute for Space Research, Groningen, The 
Netherlands and with major contributions from Germany, France and the US. 
Consortium members are: Canada: CSA, U.Waterloo; France: CESR, LAB, LERMA, 
IRAM; Germany: KOSMA, MPIfR, MPS; Ireland, NUI Maynooth; Italy: ASI, IFSI-INAF,
Osservatorio Astrofisico di Arcetri-INAF; Netherlands: SRON, TUD; Poland: CAMK,
CBK; Spain: Observatorio Astronómico Nacional (IGN), Centro de Astrobiología
(CSIC-INTA). Sweden: Chalmers University of Technology - MC2, RSS \& 
GARD; Onsala Space Observatory; Swedish National Space Board, Stockholm 
University - Stockholm Observatory; Switzerland: ETH Zurich, FHNW; USA: 
Caltech, JPL, NHSC.

PACS has been developed by a consortium of institutes led by MPE (Germany) 
and including UVIE (Austria); KU Leuven, CSL, IMEC (Belgium); CEA, LAM 
(France); MPIA (Germany); INAF-IFSI/OAA/OAP/OAT, LENS, SISSA (Italy); IAC 
(Spain). This development has been supported by the funding agencies BMVIT 
(Austria), ESA-PRODEX (Belgium), CEA/CNES (France), DLR (Germany), ASI/INAF 
(Italy), and CICYT/MCYT (Spain).

SPIRE has been developed by a consortium of institutes led by Cardiff 
University (UK) and including Univ. Lethbridge (Canada); NAOC (China); 
CEA, LAM (France); IFSI, Univ. Padua (Italy); IAC (Spain); Stockholm 
Observatory (Sweden); Imperial College London, RAL, UCL-MSSL, UKATC, 
Univ. Sussex (UK); and Caltech, JPL, NHSC, Univ. Colorado (USA). This 
development has been supported by national funding agencies: CSA 
(Canada); NAOC (China); CEA, CNES, CNRS (France); ASI (Italy); MCINN 
(Spain); SNSB (Sweden); STFC and UKSA (UK); and NASA (USA).

\end{acknowledgements}
\bibliographystyle{aa}

\begin{thebibliography}{74}
\expandafter\ifx\csname natexlab\endcsname\relax\def\natexlab#1{#1}\fi

\bibitem[{{Anders} \& {Grevesse}(1989)}]{anders89}
{Anders}, E. \& {Grevesse}, N. 1989, \gca, 53, 197

\bibitem[{{Asplund} {et~al.}(2009){Asplund}, {Grevesse}, {Sauval}, \&
  {Scott}}]{asplund09}
{Asplund}, M., {Grevesse}, N., {Sauval}, A.~J., \& {Scott}, P. 2009, \araa, 47,
  481

\bibitem[{{Boothroyd} {et~al.}(1993){Boothroyd}, {Sackmann}, \&
  {Ahern}}]{boothroyd93}
{Boothroyd}, A.~I., {Sackmann}, I.-J., \& {Ahern}, S.~C. 1993, \apj, 416, 762

\bibitem[{{Bujarrabal} {et~al.}(1994){Bujarrabal}, {Fuente}, \&
  {Omont}}]{bujarrabal94}
{Bujarrabal}, V., {Fuente}, A., \& {Omont}, A. 1994, \aap, 285, 247

\bibitem[{{Cami} {et~al.}(1998){Cami}, {de Jong}, {Justtannont}, {Yamamura}, \&
  {Waters}}]{cami98}
{Cami}, J., {de Jong}, T., {Justtannont}, K., {Yamamura}, I., \& {Waters},
  L.~B.~F.~M. 1998, \apss, 255, 339

\bibitem[{{Casassus} {et~al.}(2005){Casassus}, {Stahl}, \&
  {Wilson}}]{casassus05}
{Casassus}, S., {Stahl}, O., \& {Wilson}, T.~L. 2005, \aap, 441, 181

\bibitem[{{Chesneau} {et~al.}(2005){Chesneau}, {Verhoelst}, {Lopez}, {Waters},
  {Leinert}, {Jaffe}, {K{\"o}hler}, {de Koter}, \& {Dijkstra}}]{chesneau05}
{Chesneau}, O., {Verhoelst}, T., {Lopez}, B., {et~al.} 2005, \aap, 435, 563

\bibitem[{{Cox} {et~al.}(2012){Cox}, {Kerschbaum}, {van Marle}, {Decin},
  {Ladjal}, {Mayer}, {Groenewegen}, {van Eck}, {Royer}, {Ottensamer}, {Ueta},
  {Jorissen}, {Mecina}, {Meliani}, {Luntzer}, {Blommaert}, {Posch},
  {Vandenbussche}, \& {Waelkens}}]{cox12}
{Cox}, N.~L.~J., {Kerschbaum}, F., {van Marle}, A.-J., {et~al.} 2012, \aap,
  537, A35

\bibitem[{{Danilovich} {et~al.}(2013){Danilovich}, {Bergman}, {Justtanont},
  {Maercker}, {Olofsson}, \& {Ramstedt}}]{danilovich13}
{Danilovich}, T., {Bergman}, P., {Justtanont}, K., {et~al.} 2013, \aap, in
  preparation

\bibitem[{{De Beck} {et~al.}(2010){De Beck}, {Decin}, {de Koter}, {Justtanont},
  {Verhoelst}, {Kemper}, \& {Menten}}]{debeck10}
{De Beck}, E., {Decin}, L., {de Koter}, A., {et~al.} 2010, \aap, 523, A18

\bibitem[{{de Graauw} {et~al.}(2010){de Graauw}, {Helmich}, {Phillips},
  {Stutzki}, {Caux}, {Whyborn}, {Dieleman}, {Roelfsema}, {Aarts}, {Assendorp},
  {Bachiller}, {Baechtold}, {Barcia}, {Beintema}, {Belitsky}, {Benz}, {Bieber},
  {Boogert}, {Borys}, {Bumble}, {Ca{\"i}s}, {Caris}, {Cerulli-Irelli},
  {Chattopadhyay}, {Cherednichenko}, {Ciechanowicz}, {Coeur-Joly}, {Comito},
  {Cros}, {de Jonge}, {de Lange}, {Delforges}, {Delorme}, {den Boggende},
  {Desbat}, {Diez-Gonz{\'a}lez}, {di Giorgio}, {Dubbeldam}, {Edwards},
  {Eggens}, {Erickson}, {Evers}, {Fich}, {Finn}, {Franke}, {Gaier}, {Gal},
  {Gao}, {Gallego}, {Gauffre}, {Gill}, {Glenz}, {Golstein}, {Goulooze},
  {Gunsing}, {G{\"u}sten}, {Hartogh}, {Hatch}, {Higgins}, {Honingh}, {Huisman},
  {Jackson}, {Jacobs}, {Jacobs}, {Jarchow}, {Javadi}, {Jellema}, {Justen},
  {Karpov}, {Kasemann}, {Kawamura}, {Keizer}, {Kester}, {Klapwijk}, {Klein},
  {Kollberg}, {Kooi}, {Kooiman}, {Kopf}, {Krause}, {Krieg}, {Kramer},
  {Kruizenga}, {Kuhn}, {Laauwen}, {Lai}, {Larsson}, {Leduc}, {Leinz}, {Lin},
  {Liseau}, {Liu}, {Loose}, {L{\'o}pez-Fernandez}, {Lord}, {Luinge}, {Marston},
  {Mart{\'{\i}}n-Pintado}, {Maestrini}, {Maiwald}, {McCoey}, {Mehdi}, {Megej},
  {Melchior}, {Meinsma}, {Merkel}, {Michalska}, {Monstein}, {Moratschke},
  {Morris}, {Muller}, {Murphy}, {Naber}, {Natale}, {Nowosielski}, {Nuzzolo},
  {Olberg}, {Olbrich}, {Orfei}, {Orleanski}, {Ossenkopf}, {Peacock}, {Pearson},
  {Peron}, {Phillip-May}, {Piazzo}, {Planesas}, {Rataj}, {Ravera}, {Risacher},
  {Salez}, {Samoska}, {Saraceno}, {Schieder}, {Schlecht}, {Schl{\"o}der},
  {Schm{\"u}lling}, {Schultz}, {Schuster}, {Siebertz}, {Smit}, {Szczerba},
  {Shipman}, {Steinmetz}, {Stern}, {Stokroos}, {Teipen}, {Teyssier}, {Tils},
  {Trappe}, {van Baaren}, {van Leeuwen}, {van de Stadt}, {Visser}, {Wildeman},
  {Wafelbakker}, {Ward}, {Wesselius}, {Wild}, {Wulff}, {Wunsch}, {Tielens},
  {Zaal}, {Zirath}, {Zmuidzinas}, \& {Zwart}}]{degraauw10}
{de Graauw}, T., {Helmich}, F.~P., {Phillips}, T.~G., {et~al.} 2010, \aap, 518,
  L6

\bibitem[{{Decin} {et~al.}(2007){Decin}, {Hony}, {de Koter}, {Molenberghs},
  {Dehaes}, \& {Markwick-Kemper}}]{decin07}
{Decin}, L., {Hony}, S., {de Koter}, A., {et~al.} 2007, \aap, 475, 233

\bibitem[{{Delfosse} {et~al.}(1997){Delfosse}, {Kahane}, \&
  {Forveille}}]{delfosse97}
{Delfosse}, X., {Kahane}, C., \& {Forveille}, T. 1997, \aap, 320, 249

\bibitem[{{Dijkstra} {et~al.}(2003){Dijkstra}, {Dominik}, {Hoogzaad}, {de
  Koter}, \& {Min}}]{dijkstra03}
{Dijkstra}, C., {Dominik}, C., {Hoogzaad}, S.~N., {de Koter}, A., \& {Min}, M.
  2003, \aap, 401, 599

\bibitem[{{Driebe} {et~al.}(2005){Driebe}, {Riechers}, {Balega}, {Hofmann},
  {Men'shchikov}, \& {Weigelt}}]{driebe05}
{Driebe}, T., {Riechers}, D., {Balega}, Y., {et~al.} 2005, Astronomische
  Nachrichten, 326, 648

\bibitem[{{Dyck} {et~al.}(1974){Dyck}, {Lockwood}, \& {Capps}}]{dyck74}
{Dyck}, H.~M., {Lockwood}, G.~W., \& {Capps}, R.~W. 1974, \apj, 189, 89

\bibitem[{{Epchtein} {et~al.}(1980){Epchtein}, {Guibert}, {Nguyen-Quang-Rieu},
  {Turon}, \& {Wamsteker}}]{epchtein80}
{Epchtein}, N., {Guibert}, J., {Nguyen-Quang-Rieu}, {Turon}, P., \&
  {Wamsteker}, W. 1980, \aap, 85, L1

\bibitem[{{Fong} {et~al.}(2002){Fong}, {Justtanont}, {Meixner}, \&
  {Campbell}}]{fong02}
{Fong}, D., {Justtanont}, K., {Meixner}, M., \& {Campbell}, M.~T. 2002, \aap,
  396, 581

\bibitem[{{Garcia-Lario} {et~al.}(1997){Garcia-Lario}, {Manchado}, {Pych}, \&
  {Pottasch}}]{garcia97}
{Garcia-Lario}, P., {Manchado}, A., {Pych}, W., \& {Pottasch}, S.~R. 1997,
  \aaps, 126, 479

\bibitem[{{Goldreich} \& {Scoville}(1976)}]{goldreich76}
{Goldreich}, P. \& {Scoville}, N. 1976, \apj, 205, 144

\bibitem[{{Griffin} {et~al.}(2010){Griffin}, {Abergel}, {Abreu}, {Ade},
  {Andr{\'e}}, {Augueres}, {Babbedge}, {Bae}, {Baillie}, {Baluteau}, {Barlow},
  {Bendo}, {Benielli}, {Bock}, {Bonhomme}, {Brisbin}, {Brockley-Blatt},
  {Caldwell}, {Cara}, {Castro-Rodriguez}, {Cerulli}, {Chanial}, {Chen},
  {Clark}, {Clements}, {Clerc}, {Coker}, {Communal}, {Conversi}, {Cox},
  {Crumb}, {Cunningham}, {Daly}, {Davis}, {de Antoni}, {Delderfield}, {Devin},
  {di Giorgio}, {Didschuns}, {Dohlen}, {Donati}, {Dowell}, {Dowell}, {Duband},
  {Dumaye}, {Emery}, {Ferlet}, {Ferrand}, {Fontignie}, {Fox}, {Franceschini},
  {Frerking}, {Fulton}, {Garcia}, {Gastaud}, {Gear}, {Glenn}, {Goizel},
  {Griffin}, {Grundy}, {Guest}, {Guillemet}, {Hargrave}, {Harwit}, {Hastings},
  {Hatziminaoglou}, {Herman}, {Hinde}, {Hristov}, {Huang}, {Imhof}, {Isaak},
  {Israelsson}, {Ivison}, {Jennings}, {Kiernan}, {King}, {Lange}, {Latter},
  {Laurent}, {Laurent}, {Leeks}, {Lellouch}, {Levenson}, {Li}, {Li},
  {Lilienthal}, {Lim}, {Liu}, {Lu}, {Madden}, {Mainetti}, {Marliani}, {McKay},
  {Mercier}, {Molinari}, {Morris}, {Moseley}, {Mulder}, {Mur}, {Naylor},
  {Nguyen}, {O'Halloran}, {Oliver}, {Olofsson}, {Olofsson}, {Orfei}, {Page},
  {Pain}, {Panuzzo}, {Papageorgiou}, {Parks}, {Parr-Burman}, {Pearce},
  {Pearson}, {P{\'e}rez-Fournon}, {Pinsard}, {Pisano}, {Podosek}, {Pohlen},
  {Polehampton}, {Pouliquen}, {Rigopoulou}, {Rizzo}, {Roseboom}, {Roussel},
  {Rowan-Robinson}, {Rownd}, {Saraceno}, {Sauvage}, {Savage}, {Savini},
  {Sawyer}, {Scharmberg}, {Schmitt}, {Schneider}, {Schulz}, {Schwartz},
  {Shafer}, {Shupe}, {Sibthorpe}, {Sidher}, {Smith}, {Smith}, {Smith},
  {Spencer}, {Stobie}, {Sudiwala}, {Sukhatme}, {Surace}, {Stevens}, {Swinyard},
  {Trichas}, {Tourette}, {Triou}, {Tseng}, {Tucker}, {Turner}, {Vaccari},
  {Valtchanov}, {Vigroux}, {Virique}, {Voellmer}, {Walker}, {Ward}, {Waskett},
  {Weilert}, {Wesson}, {White}, {Whitehouse}, {Wilson}, {Winter}, {Woodcraft},
  {Wright}, {Xu}, {Zavagno}, {Zemcov}, {Zhang}, \& {Zonca}}]{griffin10}
{Griffin}, M.~J., {Abergel}, A., {Abreu}, A., {et~al.} 2010, \aap, 518, L3

\bibitem[{{Groenewegen} {et~al.}(2011){Groenewegen}, {Waelkens}, {Barlow},
  {Kerschbaum}, {Garcia-Lario}, {Cernicharo}, {Blommaert}, {Bouwman}, {Cohen},
  {Cox}, {Decin}, {Exter}, {Gear}, {Gomez}, {Hargrave}, {Henning},
  {Hutsem{\'e}kers}, {Ivison}, {Jorissen}, {Krause}, {Ladjal}, {Leeks}, {Lim},
  {Matsuura}, {Naz{\'e}}, {Olofsson}, {Ottensamer}, {Polehampton}, {Posch},
  {Rauw}, {Royer}, {Sibthorpe}, {Swinyard}, {Ueta}, {Vamvatira-Nakou},
  {Vandenbussche}, {van de Steene}, {van Eck}, {van Hoof}, {van Winckel},
  {Verdugo}, \& {Wesson}}]{groen11}
{Groenewegen}, M.~A.~T., {Waelkens}, C., {Barlow}, M.~J., {et~al.} 2011, \aap,
  526, A162

\bibitem[{{Haisch}(1979)}]{haisch79}
{Haisch}, B.~M. 1979, \aap, 72, 161

\bibitem[{{Henkel} {et~al.}(1994){Henkel}, {Wilson}, {Langer}, {Chin}, \&
  {Mauersberger}}]{henkel94}
{Henkel}, C., {Wilson}, T.~L., {Langer}, N., {Chin}, Y.-N., \& {Mauersberger},
  R. 1994, in Lecture Notes in Physics, Berlin Springer Verlag, Vol. 439, The
  Structure and Content of Molecular Clouds, ed. T.~L. {Wilson} \& K.~J.
  {Johnston}, 72

\bibitem[{{Heske} {et~al.}(1990){Heske}, {Forveille}, {Omont}, {van der Veen},
  \& {Habing}}]{heske90}
{Heske}, A., {Forveille}, T., {Omont}, A., {van der Veen}, W.~E.~C.~J., \&
  {Habing}, H.~J. 1990, \aap, 239, 173

\bibitem[{{Hofmann} {et~al.}(2001){Hofmann}, {Balega}, {Bl{\"o}cker}, \&
  {Weigelt}}]{hofmann01}
{Hofmann}, K.-H., {Balega}, Y., {Bl{\"o}cker}, T., \& {Weigelt}, G. 2001, \aap,
  379, 529

\bibitem[{{Hudgins} {et~al.}(1993){Hudgins}, {Sandford}, {Allamandola}, \&
  {Tielens}}]{hudgins93}
{Hudgins}, D.~M., {Sandford}, S.~A., {Allamandola}, L.~J., \& {Tielens},
  A.~G.~G.~M. 1993, \apjs, 86, 713

\bibitem[{{Iben} \& {Renzini}(1983)}]{iben83}
{Iben}, Jr., I. \& {Renzini}, A. 1983, \araa, 21, 271

\bibitem[{{Inomata} {et~al.}(2007){Inomata}, {Imai}, \& {Omodaka}}]{inomata07}
{Inomata}, N., {Imai}, H., \& {Omodaka}, T. 2007, \pasj, 59, 799

\bibitem[{{Johansson} {et~al.}(1977){Johansson}, {Andersson}, {Goss}, \&
  {Winnberg}}]{johansson77}
{Johansson}, L.~E.~B., {Andersson}, C., {Goss}, W.~M., \& {Winnberg}, A. 1977,
  \aap, 54, 323

\bibitem[{{Justtanont} {et~al.}(2012){Justtanont}, {Khouri}, {Maercker},
  {Alcolea}, {Decin}, {Olofsson}, {Sch{\"o}ier}, {Bujarrabal}, {Marston},
  {Teyssier}, {Cernicharo}, {Dominik}, {de Koter}, {Melnick}, {Menten},
  {Neufeld}, {Planesas}, {Schmidt}, {Szczerba}, \& {Waters}}]{justtanont12}
{Justtanont}, K., {Khouri}, T., {Maercker}, M., {et~al.} 2012, \aap, 537, A144

\bibitem[{{Justtanont} {et~al.}(2006){Justtanont}, {Olofsson}, {Dijkstra}, \&
  {Meyer}}]{justtanont06}
{Justtanont}, K., {Olofsson}, G., {Dijkstra}, C., \& {Meyer}, A.~W. 2006, \aap,
  450, 1051

\bibitem[{{Justtanont} {et~al.}(1996){Justtanont}, {Skinner}, {Tielens},
  {Meixner}, \& {Baas}}]{justtanont96}
{Justtanont}, K., {Skinner}, C.~J., {Tielens}, A.~G.~G.~M., {Meixner}, M., \&
  {Baas}, F. 1996, \apj, 456, 337

\bibitem[{{Knapp} \& {Morris}(1985)}]{knapp85}
{Knapp}, G.~R. \& {Morris}, M. 1985, \apj, 292, 640

\bibitem[{{Kozasa} \& {Sogawa}(1999)}]{kozasa99}
{Kozasa}, T. \& {Sogawa}, H. 1999, in IAU Symposium, Vol. 191, Asymptotic Giant
  Branch Stars, ed. T.~{Le Bertre}, A.~{Lebre}, \& C.~{Waelkens}, 239

\bibitem[{{Ladjal} {et~al.}(2010){Ladjal}, {Justtanont}, {Groenewegen},
  {Blommaert}, {Waelkens}, \& {Barlow}}]{ladjal10}
{Ladjal}, D., {Justtanont}, K., {Groenewegen}, M.~A.~T., {et~al.} 2010, \aap,
  513, A53

\bibitem[{{Lambert} {et~al.}(1986){Lambert}, {Gustafsson}, {Eriksson}, \&
  {Hinkle}}]{lambert86}
{Lambert}, D.~L., {Gustafsson}, B., {Eriksson}, K., \& {Hinkle}, K.~H. 1986,
  \apjs, 62, 373

\bibitem[{{Lattanzio} {et~al.}(1996){Lattanzio}, {Frost}, {Cannon}, \&
  {Wood}}]{lattanzio96}
{Lattanzio}, J., {Frost}, C., {Cannon}, R., \& {Wood}, P.~R. 1996, \memsai, 67,
  729

\bibitem[{{Lattanzio} \& {Wood}(2003)}]{lattanzio03}
{Lattanzio}, J. \& {Wood}, P.~R. 2003, in Asymptotic giant branch stars, by
  Harm J. Habing and Hans Olofsson. Astronomy and astrophysics library, New
  York, Berlin: Springer, 2003, ed. H.~J. {Habing} \& H.~{Olofsson}, 23--104

\bibitem[{{Lepine} {et~al.}(1995){Lepine}, {Ortiz}, \& {Epchtein}}]{lepine95}
{Lepine}, J.~R.~D., {Ortiz}, R., \& {Epchtein}, N. 1995, \aap, 299, 453

\bibitem[{{Likkel}(1989)}]{likkel89}
{Likkel}, L. 1989, \apj, 344, 350

\bibitem[{{Lombaert} {et~al.}(2013){Lombaert}, {Decin}, {de Koter},
  {Blommaert}, {Royer}, {De Beck}, {de Vries}, {Khouri}, \& {Min}}]{lombaert13}
{Lombaert}, R., {Decin}, L., {de Koter}, A., {et~al.} 2013, \aap, in press

\bibitem[{{Maercker} {et~al.}(2009){Maercker}, {Sch{\"o}ier}, {Olofsson},
  {Bergman}, {Frisk}, {.~Hjalmarson}, {Justtanont}, {Kwok}, {Larsson},
  {Olberg}, \& {Sandqvist}}]{maercker09}
{Maercker}, M., {Sch{\"o}ier}, F.~L., {Olofsson}, H., {et~al.} 2009, \aap, 494,
  243

\bibitem[{{Maldoni} {et~al.}(2003){Maldoni}, {Egan}, {Smith}, {Robinson}, \&
  {Wright}}]{maldoni03}
{Maldoni}, M.~M., {Egan}, M.~P., {Smith}, R.~G., {Robinson}, G., \& {Wright},
  C.~M. 2003, \mnras, 345, 912

\bibitem[{{Matsuura} {et~al.}(2009){Matsuura}, {Barlow}, {Zijlstra},
  {Whitelock}, {Cioni}, {Groenewegen}, {Volk}, {Kemper}, {Kodama}, {Lagadec},
  {Meixner}, {Sloan}, \& {Srinivasan}}]{matsuura09}
{Matsuura}, M., {Barlow}, M.~J., {Zijlstra}, A.~A., {et~al.} 2009, \mnras, 396,
  918

\bibitem[{{Mauron} \& {Huggins}(2006)}]{mauron06}
{Mauron}, N. \& {Huggins}, P.~J. 2006, \aap, 452, 257

\bibitem[{{Milam} {et~al.}(2009){Milam}, {Woolf}, \& {Ziurys}}]{milam09}
{Milam}, S.~N., {Woolf}, N.~J., \& {Ziurys}, L.~M. 2009, \apj, 690, 837

\bibitem[{{Molster} {et~al.}(2002){Molster}, {Waters}, {Tielens}, \&
  {Barlow}}]{molster02}
{Molster}, F.~J., {Waters}, L.~B.~F.~M., {Tielens}, A.~G.~G.~M., \& {Barlow},
  M.~J. 2002, \aap, 382, 184

\bibitem[{{Ohnaka} \& {Tsuji}(1996)}]{ohnaka96}
{Ohnaka}, K. \& {Tsuji}, T. 1996, \aap, 310, 933

\bibitem[{{Ohnaka} \& {Tsuji}(1999)}]{ohnaka99}
{Ohnaka}, K. \& {Tsuji}, T. 1999, \aap, 345, 233

\bibitem[{{Olofsson} {et~al.}(1993){Olofsson}, {Eriksson}, {Gustafsson}, \&
  {Carlstrom}}]{olofsson93}
{Olofsson}, H., {Eriksson}, K., {Gustafsson}, B., \& {Carlstrom}, U. 1993,
  \apjs, 87, 267

\bibitem[{{Persi} {et~al.}(1990){Persi}, {Ferrari-Toniolo}, {Ranieri},
  {Marenzi}, \& {Shivanandan}}]{persi90}
{Persi}, P., {Ferrari-Toniolo}, M., {Ranieri}, M., {Marenzi}, A., \&
  {Shivanandan}, K. 1990, \aap, 237, 153

\bibitem[{{Pilbratt} {et~al.}(2010){Pilbratt}, {Riedinger}, {Passvogel},
  {Crone}, {Doyle}, {Gageur}, {Heras}, {Jewell}, {Metcalfe}, {Ott}, \&
  {Schmidt}}]{pilbratt10}
{Pilbratt}, G.~L., {Riedinger}, J.~R., {Passvogel}, T., {et~al.} 2010, \aap,
  518, L1

\bibitem[{{Poglitsch} {et~al.}(2010){Poglitsch}, {Waelkens}, {Geis},
  {Feuchtgruber}, {Vandenbussche}, {Rodriguez}, {Krause}, {Renotte}, {van
  Hoof}, {Saraceno}, {Cepa}, {Kerschbaum}, {Agn{\`e}se}, {Ali}, {Altieri},
  {Andreani}, {Augueres}, {Balog}, {Barl}, {Bauer}, {Belbachir}, {Benedettini},
  {Billot}, {Boulade}, {Bischof}, {Blommaert}, {Callut}, {Cara}, {Cerulli},
  {Cesarsky}, {Contursi}, {Creten}, {De Meester}, {Doublier}, {Doumayrou},
  {Duband}, {Exter}, {Genzel}, {Gillis}, {Gr{\"o}zinger}, {Henning},
  {Herreros}, {Huygen}, {Inguscio}, {Jakob}, {Jamar}, {Jean}, {de Jong},
  {Katterloher}, {Kiss}, {Klaas}, {Lemke}, {Lutz}, {Madden}, {Marquet},
  {Martignac}, {Mazy}, {Merken}, {Montfort}, {Morbidelli}, {M{\"u}ller},
  {Nielbock}, {Okumura}, {Orfei}, {Ottensamer}, {Pezzuto}, {Popesso},
  {Putzeys}, {Regibo}, {Reveret}, {Royer}, {Sauvage}, {Schreiber}, {Stegmaier},
  {Schmitt}, {Schubert}, {Sturm}, {Thiel}, {Tofani}, {Vavrek}, {Wetzstein},
  {Wieprecht}, \& {Wiezorrek}}]{poglitsch10}
{Poglitsch}, A., {Waelkens}, C., {Geis}, N., {et~al.} 2010, \aap, 518, L2

\bibitem[{{Roelfsema} {et~al.}(2012){Roelfsema}, {Helmich}, {Teyssier},
  {Ossenkopf}, {Morris}, {Olberg}, {Shipman}, {Risacher}, {Akyilmaz},
  {Assendorp}, {Avruch}, {Beintema}, {Biver}, {Boogert}, {Borys}, {Braine},
  {Caris}, {Caux}, {Cernicharo}, {Coeur-Joly}, {Comito}, {de Lange},
  {Delforge}, {Dieleman}, {Dubbeldam}, {de Graauw}, {Edwards}, {Fich},
  {Flederus}, {Gal}, {di Giorgio}, {Herpin}, {Higgins}, {Hoac}, {Huisman},
  {Jarchow}, {Jellema}, {de Jonge}, {Kester}, {Klein}, {Kooi}, {Kramer},
  {Laauwen}, {Larsson}, {Leinz}, {Lord}, {Lorenzani}, {Luinge}, {Marston},
  {Mart{\'{\i}}n-Pintado}, {McCoey}, {Melchior}, {Michalska}, {Moreno},
  {M{\"u}ller}, {Nowosielski}, {Okada}, {Orlea{\'n}ski}, {Phillips}, {Pearson},
  {Rabois}, {Ravera}, {Rector}, {Rengel}, {Sagawa}, {Salomons},
  {S{\'a}nchez-Su{\'a}rez}, {Schieder}, {Schl{\"o}der}, {Schm{\"u}lling},
  {Soldati}, {Stutzki}, {Thomas}, {Tielens}, {Vastel}, {Wildeman}, {Xie},
  {Xilouris}, {Wafelbakker}, {Whyborn}, {Zaal}, {Bell}, {Bjerkeli}, {De Beck},
  {Cavali{\'e}}, {Crockett}, {Hily-Blant}, {Kama}, {Kaminski}, {Lefl{\'o}ch},
  {Lombaert}, {de Luca}, {Makai}, {Marseille}, {Nagy}, {Pacheco}, {van der
  Wiel}, {Wang}, \& {Y{\i}ld{\i}z}}]{roelfsema12}
{Roelfsema}, P.~R., {Helmich}, F.~P., {Teyssier}, D., {et~al.} 2012, \aap, 537,
  A17

\bibitem[{{Sackmann} \& {Boothroyd}(1992)}]{sackmann92}
{Sackmann}, I.-J. \& {Boothroyd}, A.~I. 1992, \apjl, 392, L71

\bibitem[{{Schoenberg} \& {Hempe}(1986)}]{schonberg86}
{Schoenberg}, K. \& {Hempe}, K. 1986, \aap, 163, 151

\bibitem[{{Sch{\"o}ier} {et~al.}(2011){Sch{\"o}ier}, {Maercker}, {Justtanont},
  {Olofsson}, {Black}, {Decin}, {de Koter}, \& {Waters}}]{schoier11}
{Sch{\"o}ier}, F.~L., {Maercker}, M., {Justtanont}, K., {et~al.} 2011, \aap,
  530, A83

\bibitem[{{Sch{\"o}ier} \& {Olofsson}(2000)}]{schoier00}
{Sch{\"o}ier}, F.~L. \& {Olofsson}, H. 2000, \aap, 359, 586

\bibitem[{{Scott} {et~al.}(2006){Scott}, {Asplund}, {Grevesse}, \&
  {Sauval}}]{scott06}
{Scott}, P.~C., {Asplund}, M., {Grevesse}, N., \& {Sauval}, A.~J. 2006, \aap,
  456, 675

\bibitem[{{Sevenster}(2002)}]{sevenster02}
{Sevenster}, M.~N. 2002, \aj, 123, 2772

\bibitem[{{Smith} \& {Lambert}(1990)}]{smith90}
{Smith}, V.~V. \& {Lambert}, D.~L. 1990, \apjl, 361, L69

\bibitem[{{Suh}(2002)}]{suh02}
{Suh}, K.-W. 2002, \mnras, 332, 513

\bibitem[{{Suh} \& {Kwon}(2009)}]{suh09}
{Suh}, K.-W. \& {Kwon}, Y.-J. 2009, Journal of Korean Astronomical Society, 42,
  81

\bibitem[{{Swinyard} {et~al.}(2010){Swinyard}, {Ade}, {Baluteau}, {Aussel},
  {Barlow}, {Bendo}, {Benielli}, {Bock}, {Brisbin}, {Conley}, {Conversi},
  {Dowell}, {Dowell}, {Ferlet}, {Fulton}, {Glenn}, {Glauser}, {Griffin},
  {Griffin}, {Guest}, {Imhof}, {Isaak}, {Jones}, {King}, {Leeks}, {Levenson},
  {Lim}, {Lu}, {Makiwa}, {Naylor}, {Nguyen}, {Oliver}, {Panuzzo},
  {Papageorgiou}, {Pearson}, {Pohlen}, {Polehampton}, {Pouliquen},
  {Rigopoulou}, {Ronayette}, {Roussel}, {Rykala}, {Savini}, {Schulz},
  {Schwartz}, {Shupe}, {Sibthorpe}, {Sidher}, {Smith}, {Spencer}, {Trichas},
  {Triou}, {Valtchanov}, {Wesson}, {Woodcraft}, {Xu}, {Zemcov}, \&
  {Zhang}}]{swinyard10}
{Swinyard}, B.~M., {Ade}, P., {Baluteau}, J.-P., {et~al.} 2010, \aap, 518, L4

\bibitem[{{Sylvester} {et~al.}(1999){Sylvester}, {Kemper}, {Barlow}, {de Jong},
  {Waters}, {Tielens}, \& {Omont}}]{sylvester99}
{Sylvester}, R.~J., {Kemper}, F., {Barlow}, M.~J., {et~al.} 1999, \aap, 352,
  587

\bibitem[{{te Lintel Hekkert} {et~al.}(1989){te Lintel Hekkert},
  {Versteege-Hensel}, {Habing}, \& {Wiertz}}]{telintel89}
{te Lintel Hekkert}, P., {Versteege-Hensel}, H.~A., {Habing}, H.~J., \&
  {Wiertz}, M. 1989, \aaps, 78, 399

\bibitem[{{Tielens} \& {Hollenbach}(1985)}]{tielens85}
{Tielens}, A.~G.~G.~M. \& {Hollenbach}, D. 1985, \apj, 291, 722

\bibitem[{{van der Veen} \& {Rugers}(1989)}]{veen89}
{van der Veen}, W.~E.~C.~J. \& {Rugers}, M. 1989, \aap, 226, 183

\bibitem[{{van Langevelde} {et~al.}(1990){van Langevelde}, {van der Heiden}, \&
  {van Schooneveld}}]{langevelde90}
{van Langevelde}, H.~J., {van der Heiden}, R., \& {van Schooneveld}, C. 1990,
  \aap, 239, 193

\bibitem[{{Vassiliadis} \& {Wood}(1993)}]{vassiliadis93}
{Vassiliadis}, E. \& {Wood}, P.~R. 1993, \apj, 413, 641

\bibitem[{{Werner} {et~al.}(1980){Werner}, {Beckwith}, {Gatley}, {Sellgren},
  {Berriman}, \& {Whiting}}]{werner80}
{Werner}, M.~W., {Beckwith}, S., {Gatley}, I., {et~al.} 1980, \apj, 239, 540

\bibitem[{{Wilson} \& {Rood}(1994)}]{wilson94}
{Wilson}, T.~L. \& {Rood}, R. 1994, \araa, 32, 191

\bibitem[{{Yuasa} {et~al.}(1999){Yuasa}, {Unno}, \& {Magono}}]{yuasa99}
{Yuasa}, M., {Unno}, W., \& {Magono}, S. 1999, \pasj, 51, 197

\end{thebibliography}

\end{document}